\documentclass{iau}
\usepackage{graphicx}
\usepackage{natbib}
\usepackage{aas_macros}

\title[AGN types and unification model] %% give here short title %%
{AGN types and unification model}

\author[Spinoglio \& Fern\'andez-Ontiveros]   %% give here short author list %%
{Luigi Spinoglio %$^1$
\and Juan Antonio Fern\'andez-Ontiveros}%$^1$}

\affiliation{%$^1$
Istituto di Astrofisica e Planetologia Spaziali - INAF, Rome, \\ Via Fosso del Cavaliere 100, 00133, Roma, Italia\\
email: {\tt luigi.spinoglio@iaps.inaf.it, j.a.fernandez.ontiveros@gmail.com} \\[\affilskip]
}

\pubyear{2019}
\volume{356}  %% insert here IAU Symposium No.
\jname{Nuclear Activity in Galaxies Across Cosmic Time}
\editors{P. Marziani, J. Masegosa, S.H. Negu, H. Netzer, M. Povic, P. Shastri {\&} S.B. Tessema, eds.}

\begin{document}

\maketitle

\begin{abstract}
The motivation of the ``unified model'' is to explain the main properties of the large zoo of active galactic nuclei with a single physical object. The discovery of broad permitted lines in the polarized spectrum of type 2 Seyfert galaxies in the mid 80's led to the idea of an obscuring torus, whose orientation with respect to our line of sight was the reason of the different optical spectra. However, after many years of observations with different techniques, including IR and mm interferometry, the resulting properties of the observed dust structures differ from the torus model that would be needed to explain the type 1 vs type 2 dichotomy. Moreover, in the last years, multi-frequency monitoring of active galactic nuclei has shown an increasing number of transitions from one type to the other one, which cannot be explained in terms of the simple orientation of the dusty structure surrounding the active galactic nucleus (AGN). The interrelations between the AGN and the host galaxy, as also shown in the Magorrian relation, suggest that the evolution of the host galaxy may also have an important role in the observed manifestation of the nuclei. As an example, the observed delay between the maximum star formation activity and the onset of the AGN activity, and the higher occurrence of type 2 nuclei in star forming galaxies, have suggested the possible evolutionary 
%star formation appears to be more frequent in type 2 compared to type 1 objects. This might be related to a possible evolutionary 
path from, e.g., \textsc{H\,ii} $\rightarrow$ AGN2 $\rightarrow$ AGN1. In the next years the models of unification need to also consider this observational framework and not only simple orientation effects. 
\keywords{accretion, galaxies: active, galaxies: nuclei, galaxies: Seyfert, ISM: dust, galaxies: evolution}
\end{abstract}

\firstsection % if your document starts with a section,
\section{Introduction}
Since their discovery in 1943, Seyfert galaxies appeared to show different optical spectra, as already described in the paper by \citet{sey43}: ``The observed relative intensities of the emission lines exhibit large variations from nebula to nebula''. As an example of this variety, the maximum observed widths of hydrogen recombination lines were ranging from $3600$ to $8500\, \rm{km\,s^{-1}}$. In a following paper, \citet{sey46} showed that ``the hydrogen lines in NGC\,4151 and NGC\,7469 are of unusual interest, being composed of relatively narrow cores ($1100\, \rm{km\,s^{-1}}$) superposed on very wide wings ($7500\, \rm{km\,s^{-1}}$)''.

The main difference between type 1 and type 2 Seyfert galaxies come from the optical and UV spectroscopy: \textit{i)} type 1 have broad permitted lines (i.e., hydrogen recombination lines, intercombination lines --\,e.g., \textsc{C\,iv} 1459\,\AA, \textsc{Mg\,ii} 2798\,\AA \,-- and semi-forbiden lines, e.g., \textsc{C\,iii}] 1909\,\AA) with line widths of FWHM\,$\sim 1000-10000\, \rm{km\,s^{-1}}$; \textit{ii)} both type 1 and type 2 have narrow permitted and forbidden lines: FWHM\,$\sim 500-1000\, \rm{km\,s^{-1}}$.

The complexity and variety of Active Galactic Nuclei (AGN), which extend the class of Seyfert galaxies to higher luminosities, has been acknowledged in the following decades as the observational material were accumulated. The active galactic nuclei {\it zoo} has been presented in a coherent framework in the review by \citet[][see their Fig.\,4]{hec14}: the main parameter is the total luminosity in Eddington units, which is linked to the accretion rate of the supermassive black holes (SMBH) and sets the transition from the jet mode to the radiative mode.

\begin{figure}
% \vspace*{-2.0 cm}
\begin{center}
    \includegraphics[width=3.5cm]{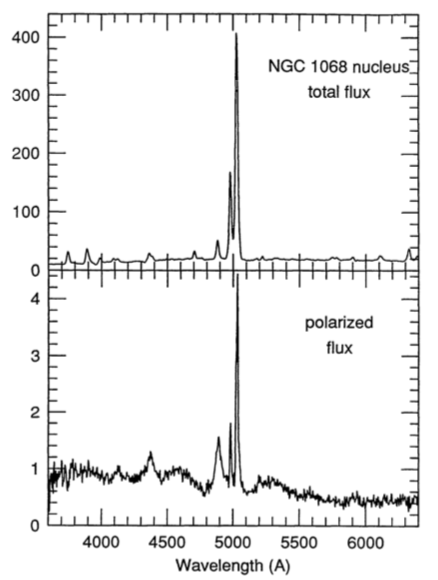}~\includegraphics[width=10.2cm]{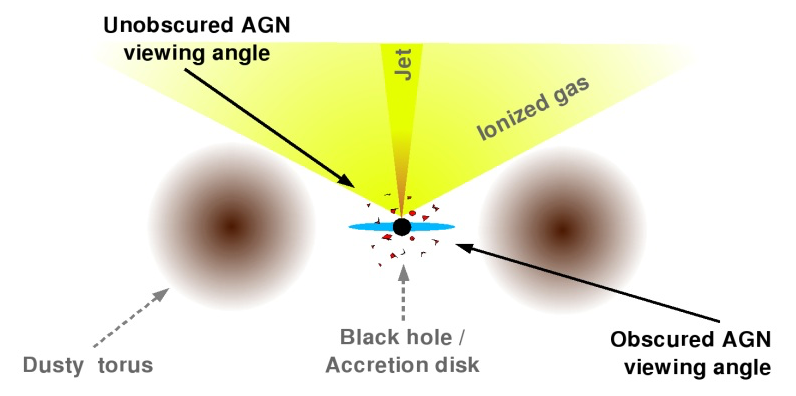}
% \vspace*{-1.0 cm}
    \caption{{\bf Left: (a)}  Spectropolarimetry of NGC~1068, showing (above) the flux spectrum and below the polarized flux, which is indistinguishable from the flux spectra of Type 1 spectra. Figure from \citet{ant93}. {\bf Right: (b)} Schematic view of the torus model. Figure from \citet{gan05}.}\label{fig_ant93}
\end{center}
\end{figure}

 In the mid '80s, \citet{ant85} first reported that the optical polarization spectrum of the prototype Type 2 Seyfert galaxy NGC\,1068 showed very broad symmetric Balmer lines ($\sim 7500\, \rm{km\,s^{-1}}$) and also permitted \textsc{Fe\,ii}, indicating a very close similarity with Type 1 Seyfert galaxies (see Fig.\,\ref{fig_ant93}-a). In their paper, they say ``We favor an interpretation in which the continuum source and broad line clouds are located inside a thick disk, with electrons above and below the disk scattering continuum and broad line photons in to the line of sight''. This paper coincides with the beginning of the ``Unification Era'', i.e. the attempt to explain the different spectral manifestations of AGN with a single physical object, viewed from different angles. In his review, \citet{ant93} introduced what he named the ``straw person model'', for which there are only two types of AGN: those radio quiet and those radio loud. Considering only radio quiet objects, this statement means that ``all properties such as spectroscopic classification [...] are ascribed to orientation'' \citep{ant93}. 

We can anticipate here that the situation, after three decades of observations, more and more evidence has shown that the simple unification scheme proposed by \citet{ant93} is far too simplistic to account for them: if on the one hand it can be recognised that some unification is well proven by observations, i.e. the so-called ``Hidden Broad-Line Region Galaxies'' (HBLR) are the same as the type 1 AGN, on the other hand, the occurrence of other intrinsic factors in the nature and appearance of AGN are present. Among these: the effect of evolution of the accretion disk, which is not a stable physical system; the effect of the host galaxy, which is not negligible. Moreover, the simple unification scheme relies on the presence of a geometrically thick obscuring structure, the so-called dusty torus, whose robust observational evidence is still not available.

This review is organized as follows: Section 2 introduces the concept of the hypothetical torus which would be needed to demonstrate the unification between type 1 and type 2 AGN; Section 3 describes some of the observational material which has been collected so far with different techniques aiming to the detection of the molecular tori around SMBHs; Section 4 describes the unification which has been already reached for low luminosity AGN; Section 5 tries to make an observational review, without any ambition to be complete, of the data on the so-called {\it changing look} AGN, i.e. those AGN which have been transited from one type to another one and may question the simple unification model. Section 6 attempts to assess the role of evolution and of the host galaxy in the AGN appearance. Finally, in Section 7 we present our conclusions.

\section{The needed molecular torus}

A large number of works have been published on models of molecular tori, either theoretical and phenomenological. A comprehensive review of what has been done up to the mid 2010's is that by \citet{net15}, to which we refer the reader to obtain a full comprehension of the various approaches used in the literature to investigate AGN tori, an essential work to assess the unification model. Other relevant reviews on obscuration around AGN and obscured AGN are those of \citet{ram17}, \citet{hic18}, and the updated review by \citet{ant12} respectively. In this review, we will limit our attention here to understand what characteristics a torus needs to have in order to be compliant with the unification model. We show, in Fig.\,\ref{fig_ant93}-b, the schematic view of the torus model from \citet{gan05}.
The expected characteristics and the physical role of the torus are:
\begin{itemize}
    \item It needs to block the Broad Line Region (BLR) radiation. This is the main requirement of the torus to allow for the main observational difference between type 1 and type 2 AGN, which is the appearance of the optical and ultraviolet emission line spectra. A strong observational constraint on this can be derived from the statistics of the fraction of type 1 versus type 2 AGN seen, e.g., in the Local Universe. This fraction, however, strongly depends on the AGN luminosity. At the typical Seyfert galaxy luminosities of ${\rm L_{bol}~{\leq}~10^{44}~erg~s^{-1}}$ the Type 1 fraction is $\sim$0.3, increasing to $\sim$0.7 at ${\rm L_{bol}~{\leq}~10^{46}~erg~s^{-1}}$ \citep{ass13}.
    Recent hard X-ray surveys have significantly improved our understanding of AGN obscuration, showing that ~70\% of all local AGN are obscured \citep{bur11,ric15}.
    If the opening angle of the hypothetical torus (shaded angle in Fig.\,\ref{fig_ant93}-b) is, e.g., 120$^{\circ}$, then the surface area with BLR visibility will intersect exactly half of the hemisphere visible from an observer, which will correspond to an equal fraction of types 1 and types 2. However, if this angle is as narrow as 60$^{\circ}$, then the fraction of type 1 should be smaller than 15\%.
    %{\bf [..XXX try to conclude something here..]}
    
    \item It needs to collimate the AGN ionizing radiation and induce the biconical shape of the Narrow Line Region (NLR), i.e the so-called ionization cones: studies on the NLR morphology are relevant in this respect.
    
    \item It needs to allow the feeding of the Black-Hole through accretion, and act at the same time as a gas reservoir.
\end{itemize}

\section{Observations of the ``torus''}\label{sec_torus}
\subsection{Mid-IR interferometry}
We describe in this Section the most relevant ``direct" observations of the torus, through interferometric techniques, either in the mid-IR and in the millimeter ranges. The first solid claim of the detection of a torus has been reported by \citet{jaf04} for the prototype Seyfert type 2 galaxy NGC\,1068. It is not surprising that this galaxy was chosen, because of the discovery of broad hydrogen lines in the polarized optical spectrum by \citet{ant85} and because of its vicinity (only $14.4\, \rm{Mpc}$ from us).

\begin{figure}[t]
% \vspace*{-2.0 cm}
\begin{center}
    \includegraphics[width=6.7cm]{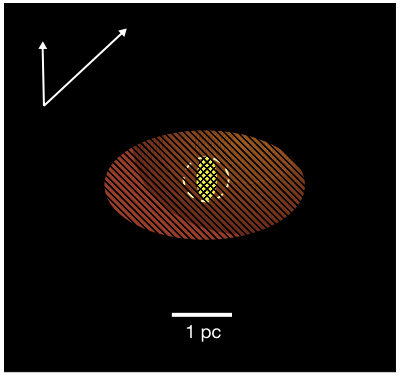}~
    \includegraphics[width=7.0cm]{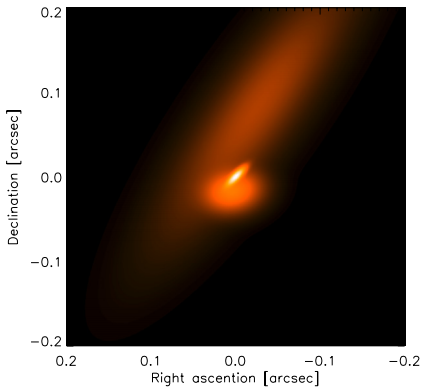}
% \vspace*{-1.0 cm}
    \caption{{\bf Left: (a)} Model dust structure in the nucleus of NGC~1068, showing a central hot component (dust temperature T $>$ 800~K, light) marginally resolved along the source axis. The much larger warm component (T = 320~K, dark shade) is well resolved. Single hatching represents the averaged optical depth in the silicate absorption, $\tau_{SiO}$ = 0.3, while the higher value $\tau_{SiO}$ = 2.1 is found towards the hot component (cross-hatched). Figure from \citet{jaf04}. {\bf Right: (b)} Image of the three component model 2 of \citet{lop14} for the mid-infrared emission at 12$\mu$m of the nuclear region of NGC~1068. Figure from \citet{lop14}.} \label{fig_jaffe04}
\end{center}
\end{figure}

\citet{jaf04} report interferometric mid-IR observations with the MIDI \citep{lei03} instrument at the focal plane of the ESO-VLTI \citep{gli03} interferometer, reaching a spatial resolution of $\sim 10\, \rm{mas}$ at $\lambda = 10\, \rm{\mu m}$, that spatially resolve the mid-IR brightness distribution in the nucleus of NGC\,1068. These observations were apparently consistent with a warm ($320\, \rm{K}$) dust structure with a size of of $2.1 \times 3.4\, \rm{pc^2}$, enclosing a smaller and hotter structure (Fig.\,\ref{fig_jaffe04}-a). However, these results were based on a relatively poor coverage or spatial frequencies in the $uv$-plane and therefore the modeling of the observed visibilities needed to reconstruct the torus geometry has to be taken with extreme caution. As a matter of fact, the $uv$-converage was improved later by \citet{lop14}, with the same instrument, but including longer baselines thanks to the Auxiliary Telescopes, allowing to measure also the $5 - 10\, \rm{pc}$ scales. Their results show that most of the mid-IR emission originates from the large-scale structures and is associated with warm dust distributed in two major components, one close to the center and one at a distance of more than $80\, \rm{mas}$, close to $16^{\circ} - 18^{\circ}$ in the NW direction (Fig.\,\ref{fig_jaffe04}-b). While the central warm region is interpreted as an extension of the hot emission region, the offset region is  attributed to dusty clouds close to the northern ionization cone. Models infer a size of $14\, \rm{pc}$, strong elongation along a position angle of $\rm{PA} \sim -35^{\circ}$, and three times more $12\, \rm{\mu m}$ flux than that of the central hot region.

%If similar regions are heated by the direct radiation from the nucleus, then they will contribute substantially to the classification of many Seyfert galaxies as Type 2. Such a region is also consistent in other Seyfert galaxies (the Circinus galaxy, NGC~3783, and NGC~424). 

The warm component, named 3 in \citet{lop14}, located $\sim 7\, \rm{pc}$ north of the hotter nuclear disk, apparently intercepts a large fraction of the nuclear UV emission. Thus there are several obscuring components at different disk latitudes that can cause type 2 appearance in AGN. The volume that is heated by this emission is quite narrow; the viewing angles from which this galaxy would be classified as Seyfert type 1 cover only $\sim 10\%$ of the sky. Even if the spatial frequency sampling of these data is very good, compared to the previous work, however the best model does not appear to have the characteristics of the needed torus.

\begin{figure}[t]
% \vspace*{-2.0 cm}
\begin{center}
    \includegraphics[width=10cm]{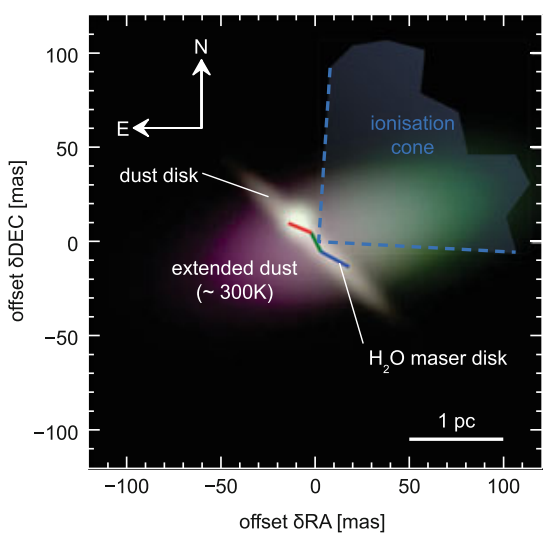}
% \vspace*{-1.0 cm}
    \caption{Image of the three-component model for the mid-IR emission of the nucleus of the Circinus galaxy. The gradient of the extended component due to the increase in the silicate depth towards the south-east is clearly visible. Despite the lower surface brightness, 80\% of the emission originates from the extended component. The trace of the water maser disk is also plotted: the blue and red parts trace the approaching and receding sides of the maser disk respectively. Figure from \citet{tri14}, adapted by \citet{hoe16}.}\label{fig_tristram14}
\end{center}
\end{figure}

\citet{tri14} collected extensive mid-IR interferometric data of the closest Seyfert type 2 galaxy Circinus (at $\sim$4 Mpc) with MIDI at the VLTI during many observing runs from 2008 to 2011. Besides confirming the presence of two components in the dust distribution, one inner dense disk component and an extended emission region, they analyse with detailed models these components. 
For a better understanding, we report their figure here (Fig.\,\ref{fig_tristram14}).

The disk-like component is highly elongated (along $\rm{PA} \simeq 46^{\circ}$) with a size of $\sim 0.2 \times 1.1\, \rm{pc^2}$, aligned with the orientation of the nuclear maser disk and perpendicular to the ionisation cone and outflow (along $\rm{PA} \sim -44^{\circ}$). They interpret this component either as emission from material associated with the inner funnel of the torus directly above or below the disk or from the directly illuminated portion of a warped disk slightly oriented towards us.

The extended dust emission accounts for $80\%$ of the mid-IR emission. It has a size of $\sim 0.8 \times 1.9\, \rm{pc^2}$ and is elongated along $\rm{PA} \sim 107^{\circ}$, that is roughly along the polar direction. It is interpreted as the emission from the inner funnel of a more extended dust distribution and especially as emission from the funnel edge along $\rm{PA} \sim -90^{\circ}$. Dense dusty material enters the ionisation cone primarily on this side of the funnel, which is also preferentially illuminated by the inclined accretion disk.

They find both emission components to be consistent with dust at $T \sim 300\, \rm{K}$, i.e. no evidence of an increase in the temperature of the dust towards the center, meaning that most of the near-IR emission probably comes from parsec scales as well. They further argue that {\it the disk component alone is not sufficient to provide the necessary obscuration and collimation of the ionising radiation and outflow}. The material responsible for this must instead be located on scales of $\sim 1\, \rm{pc}$, surrounding the disk. They associate this material with the dusty torus. However, they conclude that the presence of a bright disk-like component, polar elongated dust emission and the lack of a temperature difference are not expected for typical models of the centrally heated dust distributions of AGN. Moreover, they state that new sets of detailed radiative transfer calculations will be required to explain their observations and to better understand the three-dimensional dust morphology in the nuclei of active galaxies.

While initial observations using mid-IR interferometry initially confirmed the presence of a geometrically-thick obscuring structure in NGC\,1068 \citep{jaf04}, further observing campaigns on prototypical hidden Seyfert type 1 galaxies favor a geometrically-thin dust distribution, i.e. a disk instead of a torus. This casts doubts on the role that this structure plays on the obscuration and the type 1/type 2 dichotomy.

\subsection{Redefining the torus}

%Criticalities: 
%\begin{itemize}
%    \item How the main dusty structure can have the role of blocking BLR radiation ?
%    \item Would this be consistent with statistical arguments (fraction type 2/type 1) ?
%    \item How the inner region can be consistent with the {\it needed} torus ?
%\end{itemize}

%(H{\"o}nig 2019)

\begin{figure}
% \vspace*{-2.0 cm}
\begin{center}
    \includegraphics[width=10cm]{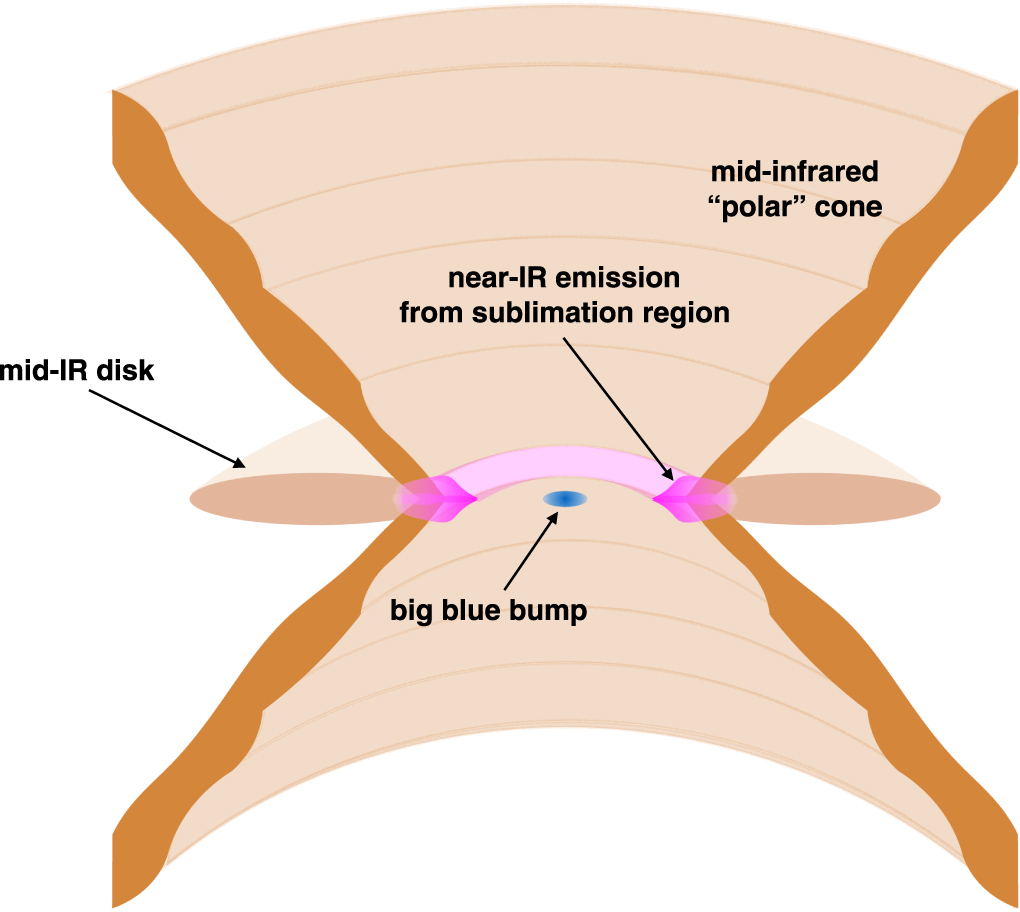}
% \vspace*{-1.0 cm}
    \caption{Schematic view of the pc-scale AGN infrared emission consisting of a geometrically-thin disk in the equatorial plane (light brown) and a hollow dusty cone towards the polar region (dark brown). The inner part of the disk (pink) emits the near-IR emission dominating the $3-5\, \rm{\mu m}$ bump. Figure from \citet{hoe19}.}\label{fig_hoenig19}
\end{center}
\end{figure}

Molecular lines show large, massive disks while mid-IR observations are dominated by a strong polar component, which is interpreted as a dusty wind.
 A unifying view of AGN in the infrared (IR) and sub-mm has been recently proposed by \citet{hoe19}. His paper aims at using characteristics shared by AGN in each of the wavebands and a set of simple physical principles to form a unifying view of these seemingly contradictory observations: 
 \begin{itemize}
\item Dusty molecular gas flows in from galactic scales of $\sim 100\, \rm{pc}$ to the sub-pc environment via a disk with small to moderate scale height.
\item The hot, inner part of the disk inflates due to IR radiation pressure and unbinds a large amount of the inflowing gas from the BH gravitational potential, providing the conditions to launch a wind driven by the radiation pressure from the AGN.
\item The dusty wind feeds back mass into the galaxy at a rate of the order of $\sim$0.1-100 ${\rm M_{\odot}\,yr^{-1}}$, depending on AGN luminosity and Eddington ratio. 
\item Angle-dependent obscuration as required by AGN unification is provided by a combination of disk, wind, and wind launching region. 
\end{itemize}

\citet{gon19a, gon19b} recently made an investigation of the predictions of six dusty torus models of AGN, including either clumpy torus models, two phase models and also the wind/torus model \citep{hoe17}, all with available spectral energy distributions, aiming at exploring which model describes better the data and the resulting parameters. They show that 
different torus models explain the same energy distributions with very different geometries, such as the viewing angle and the covering factor. Being these two parameters among the most important physical quantities characterizing the AGN appearance, we conclude that the torus models in general do not appear to be robustly linked to the observed properties, making their scope to link theory to observations very difficult.

\subsection{ALMA interferometry}

\begin{figure}[t]
% \vspace*{-2.0 cm}
\begin{center}
    \includegraphics[width=12cm]{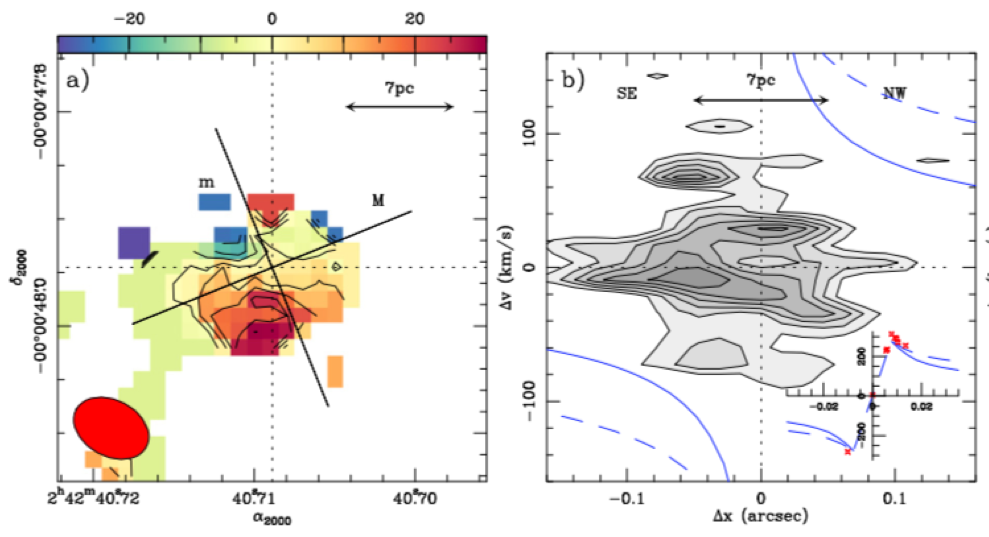}
% \vspace*{-1.0 cm}
    \caption{{\bf Left:(a)} ALMA CO(6-5) mean-velocity map of the torus. The black lines labeled as ``M'' and ``m'' show, respectively, the orientations of the major and minor axes of the CO torus. 
    {\bf Right:(b)} CO(6-5) position-velocity diagram along the M axis. The inset shows the velocities relative to $v_{\rm sys}$ as a function of radius (in arcseconds) as derived for the $\rm {H_{2}O}$ megamaser spots (red markers) detected along $\rm{PA}^{\rm maser} = 140^{\circ} \pm 5^{\circ}$ \citep{gre96}. The dashed (solid) blue curve shows the best-fit sub-Keplerian (Keplerian) rotation curve $v_{\rm rot} \propto r^{-a}$ of \citet{gre96} with $\alpha = 0.31$ ($0.50$). Figure from \citet{gar16}.} \label{fig_garcia16}
\end{center}
\end{figure}
\citet{gar16} used the Atacama Large Millimeter Array (ALMA) interferometer \citep{woo09} to map the NGC~1068 circumnuclear disk (CND), extended $\sim 300\, \rm{pc}$, in the CO(6-5) transition and in the continuum at $432\, \rm{\mu m}$  at a spatial resolution of $\sim 4\, \rm{pc}$.

The CND has been spatially resolved and its dust emission imaged, the molecular gas component shows a disk with a $7 - 10\, \rm{pc}$ diameter (Fig.\,\ref{fig_garcia16}-a). However, no clear rotation pattern is present in the data, even if this molecular disk might be a gas reservoir, it is hard to say that this is a stable structure. Moreover, the CO emission does not appear to be related to the H$_2$O masers orbiting the SMBH \citep{gre96}, indicating that the CO gas is not in Keplerian motion around the BH, as one would expect (Fig.\,\ref{fig_garcia16}-b). The dynamics of the molecular gas in the torus show instead strong non-circular motions and enhanced turbulence superposed on a surprisingly slow rotation pattern of the disk. How can we explain the BH feeding from such a turbulent structure? How universal is this picture? 

\begin{figure}
% \vspace*{-2.0 cm}
\begin{center}
    \includegraphics[width=10cm]{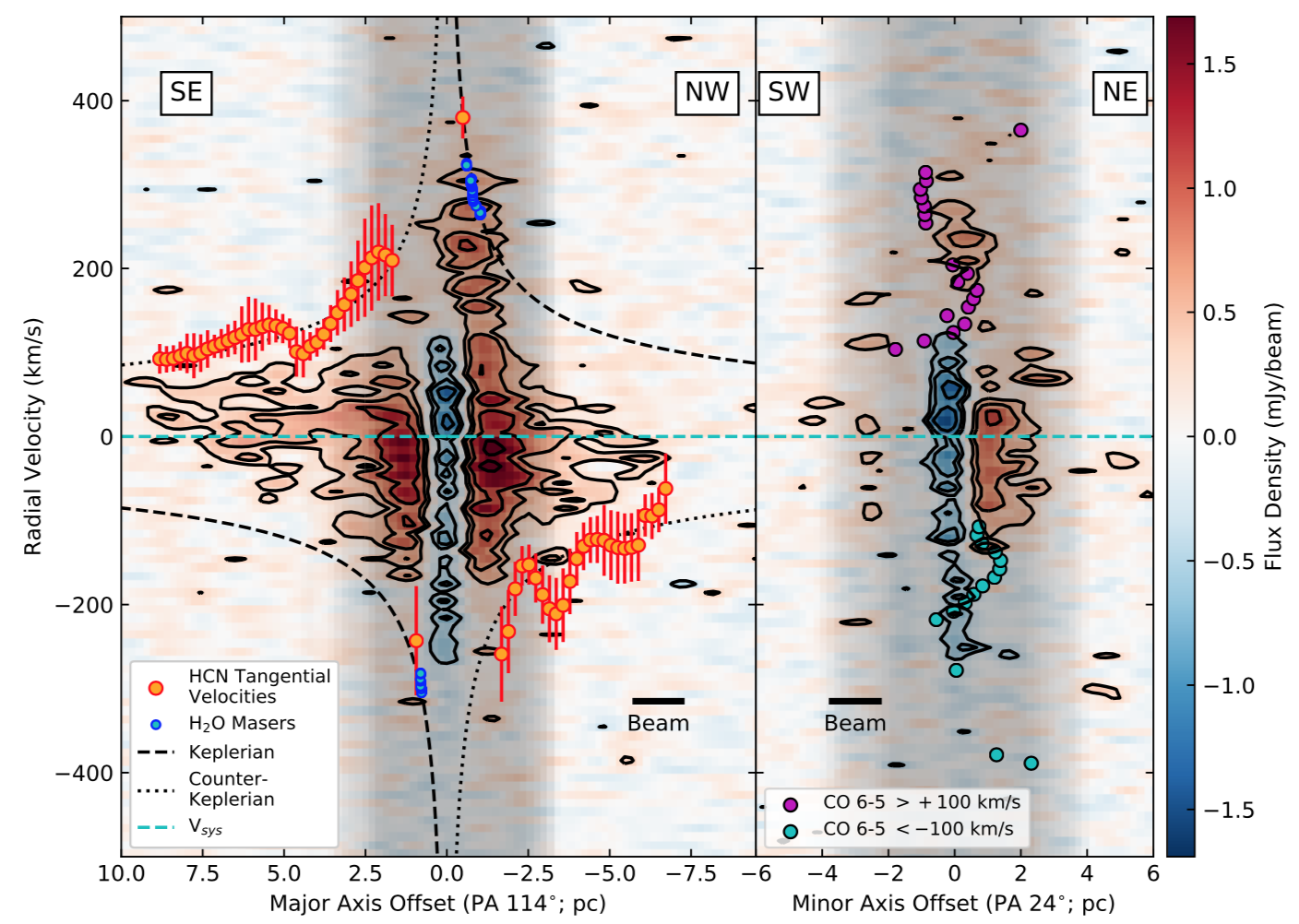}
% \vspace*{-1.0 cm}
    \caption{ALMA HCN (J = 3-2) position-velocity diagrams extracted along the HCN major axis (left panel) and minor axis (right panel) of the nuclear region of NGC\,1068.
%    The major axis offsets run southeast (positive offset) to northwest (negative offset) along PA 114$^{\circ}$ through the nuclear continuum source S1, and the minor axis offsets run southwest (negative offset) to northeast (positive offset) along PA 24$^{\circ}$.
On the major axis diagram, the tangential HCN velocities are marked by yellow dots with red errorbars, and, for comparison, the H$_{2}$O maser radial velocities along $\rm{PA} \sim 131^{\circ}$ are shown as filled blue circles (data from \citealt{gre97}) corrected for the best-fit inclination, $i = 79^{\circ}$. The best-fitting model for the H$_{2}$O masers, shown with a dashed line, is Keplerian rotation around a central mass of $1.66 \times 10^7\, \rm{M_\odot}$. %(inner disk rotation; Gallimore \& Impellizzeri 2019). 
%The dotted line shows the same Keplerian rotation, but in the opposite sense. The 256 ${\rm GHz}$ continuum emission is indicated by the shaded gray regions. 
Figure from \citet{imp19}.} \label{fig_impel19}
\end{center}
\end{figure}

\citet{imp19} mapped the inner region of tens of parsecs around the nucleus of NGC\,1068 in the HCN (J = 3-2) transition at 256 GHz. They identify three kinematically distinct regions: (1) an outflow component in emission on the HCN position-velocity diagram and detected as a blueshifted wing in absorption against the nuclear continuum source, with projected outflow speeds approaching $\sim 450\, \rm{km\,s^{-1}}$; (2) an inner disk with radius of $\sim 1.2\, \rm{pc}$; and (3) an outer disk extending to $r \sim 7\, \rm{pc}$. The two disks counter-rotate, and the kinematics of the inner disk agree with the H$_{2}$O megamaser disk mapped by the VLBA \citep{gre97}. The outer disk shows a Keplerian rotation curve consistent with an extrapolation of the rotation curve of the inner disk. They also find that the HCN radial velocity field is more complex along the molecular outflow axis, which suggests detecting but not fully resolving HCN emission associated with the outflow (Fig.\,\ref{fig_impel19}). They conclude that the molecular obscuring medium in NGC\,1068 consists of counter-rotating and misaligned disks on parsec scales. It appears therefore that the observational situation is far more complicate than we would expect from the {\it simple} torus.

\section{Low Luminosity AGN Unification}

\begin{figure}[t]
\begin{center}
% \vspace*{-2.0 cm}
    \includegraphics[width=14cm]{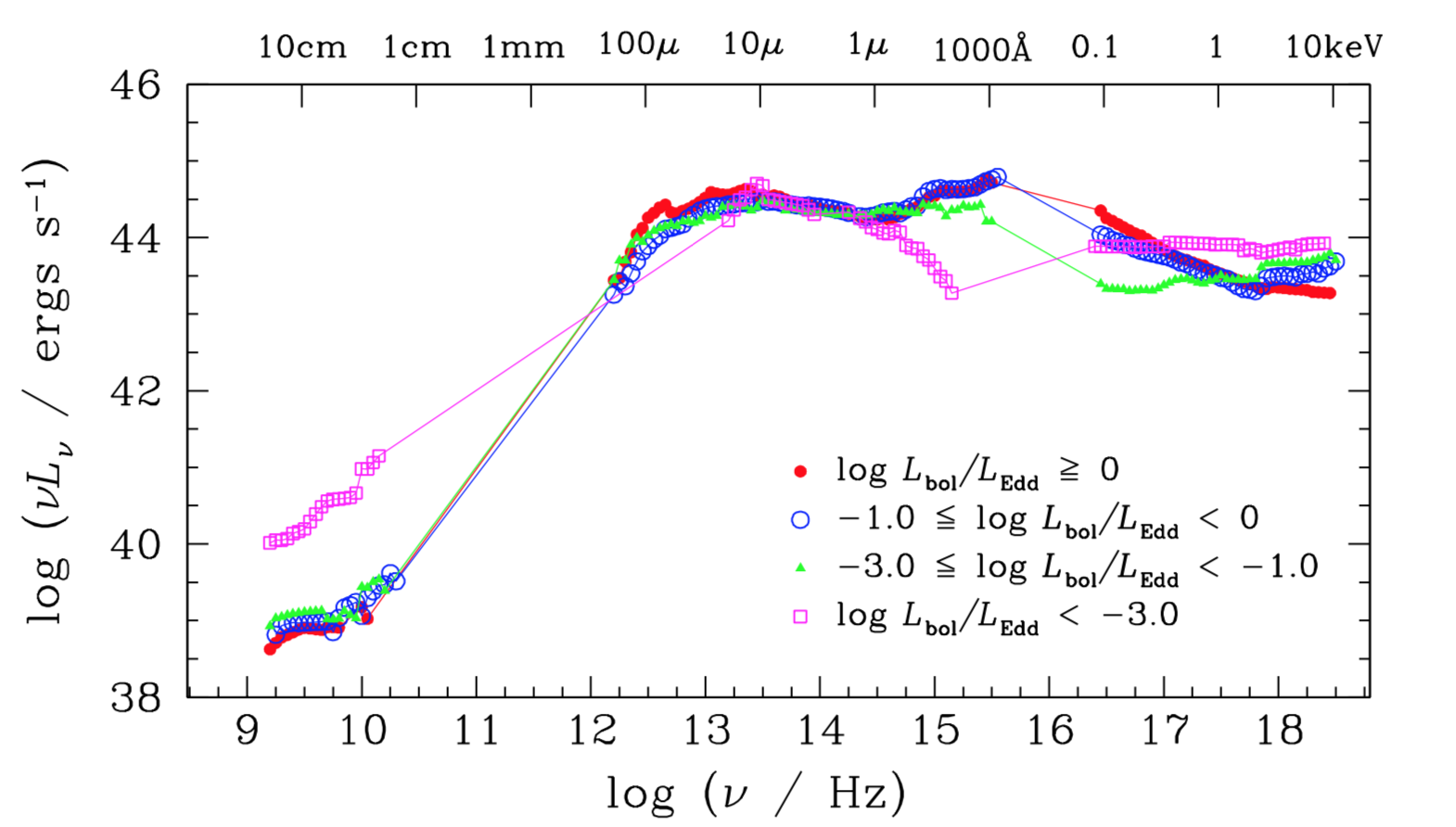}
% \vspace*{-1.0 cm}
\caption{Composite SEDs for radio-quiet AGN binned by Eddington ratio. The SEDs are normalized at $1\, \rm{\mu m}$. Figure from \citet{ho08}.}\label{fig_ho08}
\end{center}
\end{figure}

At low luminosities, AGN seem to depart from the classical unification scheme. The big blue bump --\,associated to the thermal emission from the accretion disk\,-- is typically missing in these nuclei (Fig.\,\ref{fig_ho08}), whose emission is usually dominated by a featureless, non-thermal power-law continuum \citep{ho96,ho08}. Furthermore, some models predict the vanishing of the torus at low AGN luminosities ($\log(L_{\rm bol}/L_{\rm edd}) \lesssim -3$; $L_{\rm bol} \lesssim 10^{42}\, \rm{erg\,s^{-1}}$), when the radiation pressure from the nucleus can no longer sustain this structure \citep{eli06}, following also the collapse of the BLR \citep{nic03,eli09}. Radiatively inefficient accretion flow models \citep{nar95b} and jet outflow models \citep{fal00} were introduced to explain the disappearance of the accretion disk and ultimately the low radiative efficiency observed in a variety of sources across a wide range in BH masses, from X-ray binaries in the low-hard state \citep{mar01} to Sgr A$^*$ \citep{nar95a,fal00} and Low Luminosity AGN \citep[LLAGN][]{yu11,mar08}. In these models the accretion disk recedes at low accretion rates, and the innermost region is dominated by a geometrically-thick structure where the material is either advected towards the BH or ejected along the system axis forming a collimated jet or wind outflow. This causes the disappearance of the big blue bump, since viscous phenomena are not acting within the innermost and hottest disk radii, and the advected material does not radiate the energy gained through accretion. On the other hand, the higher radio activity observed in LLAGN is associated to the jet \citep{ho08}.

\begin{figure}[t]
% \vspace*{-2.0 cm}
\begin{center}
    \includegraphics[width=14cm]{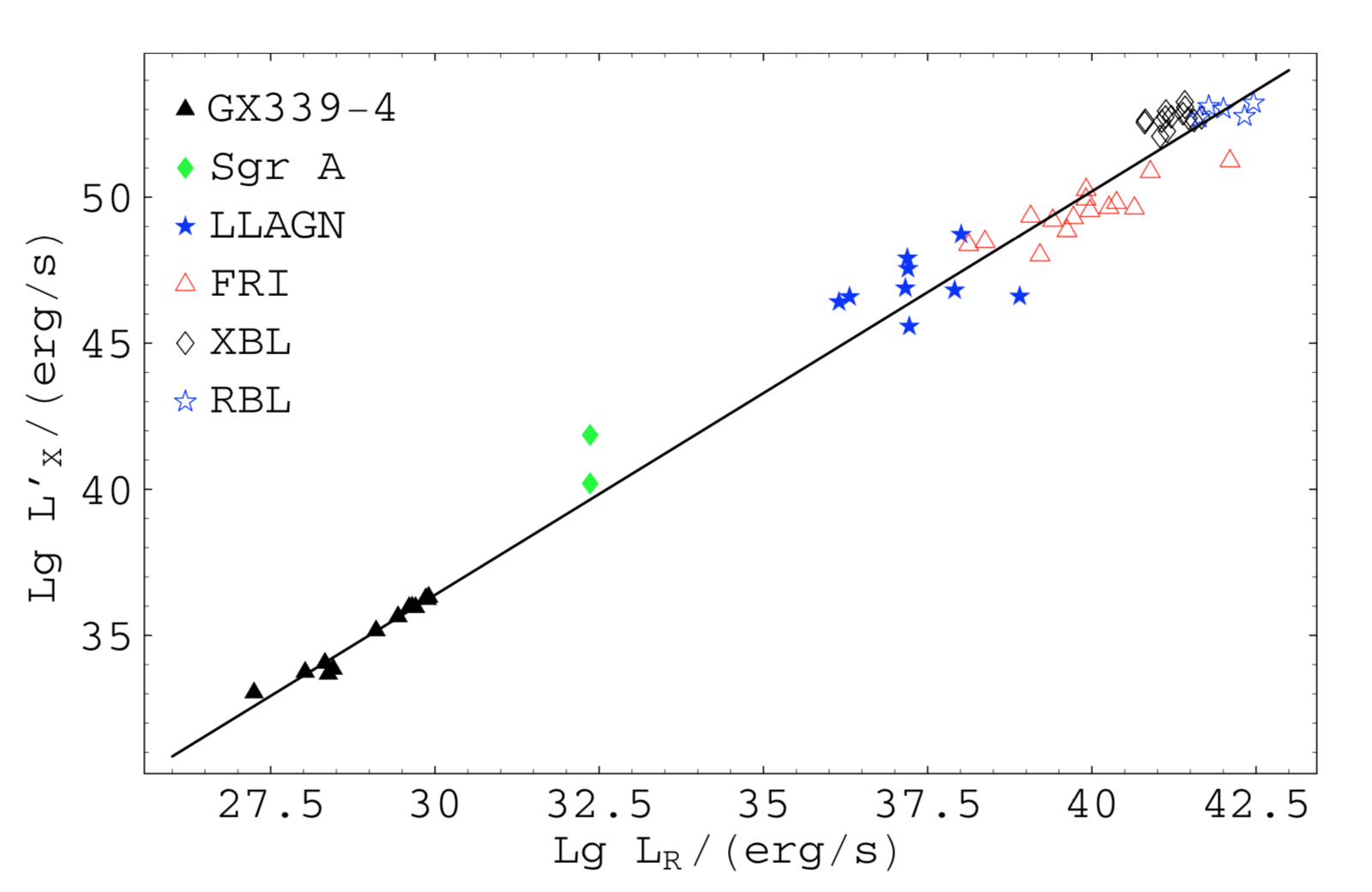}
% \vspace*{-1.0 cm}
    \caption{Radio/X-ray correlation for XRBs and AGN, where the X-ray flux of all AGN has been increased by a constant value of $10^7$, corresponding to an average AGN mass of $3 \times 10^9\, \rm{M_\odot}$. Figure from \citet{fal04}.}\label{fig_fal04}
\end{center}
\end{figure}

This common framework to explain the physics of accretion at low luminosities have provided a wide unification for low-power BHs across the mass spectrum, which was confirmed by the discovery of the fundamental plane of BH accretion \citep[][Fig.\,\ref{fig_fal04}]{mer03,fal04}. The latter is a tight correlation found between the continuum emission at radio, the X-ray flux, and the BH mass, from stellar-mass BH to SMBHs, proving the common physics of the low-hard state in these systems over $\sim 8$ orders of magnitude in mass.

\section{The Mutant AGN}\label{mutants}

\begin{figure}
% \vspace*{-2.0 cm}
\begin{center}
    \includegraphics[width=7.5cm]{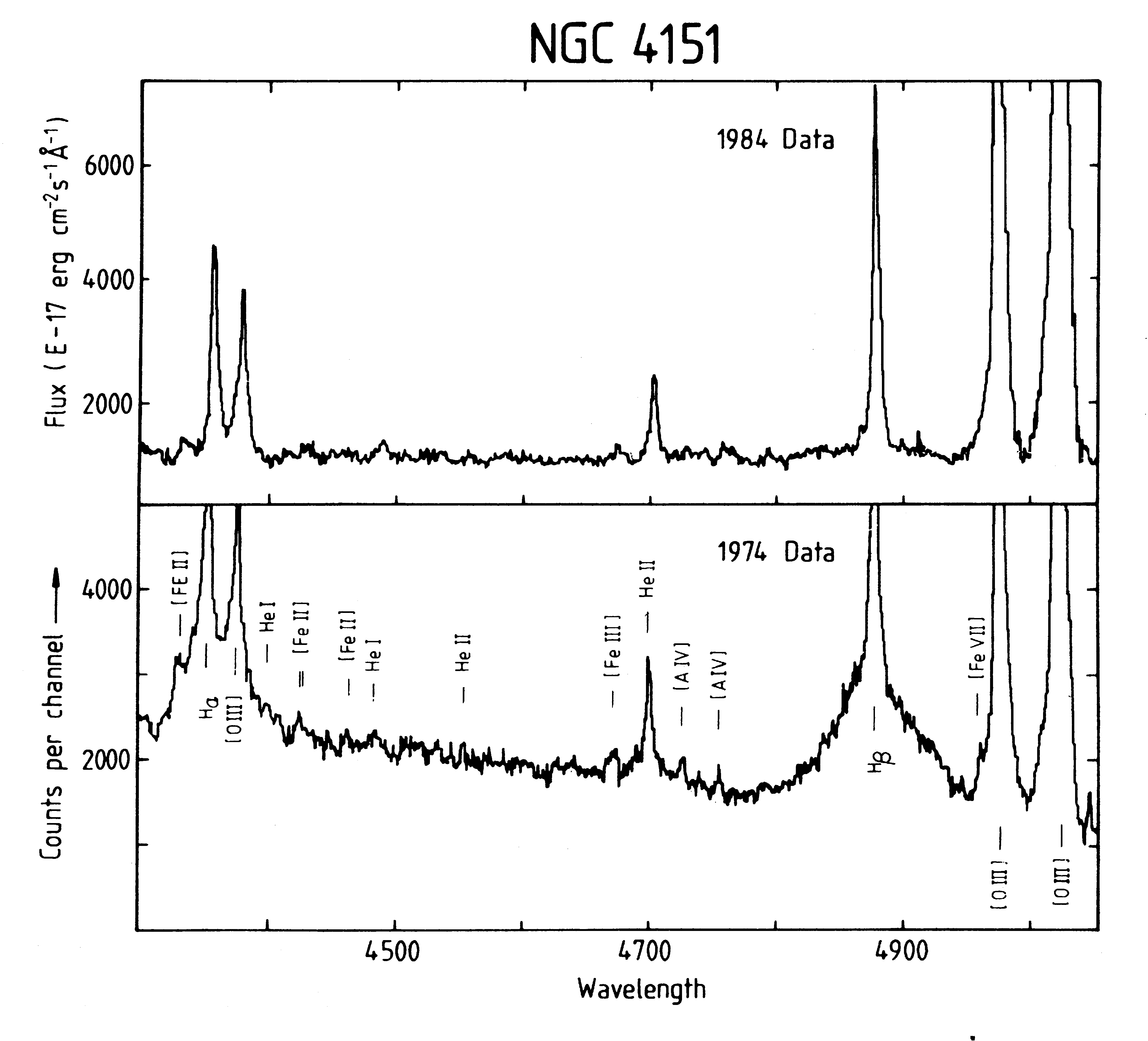}~
    \includegraphics[width=5.8cm]{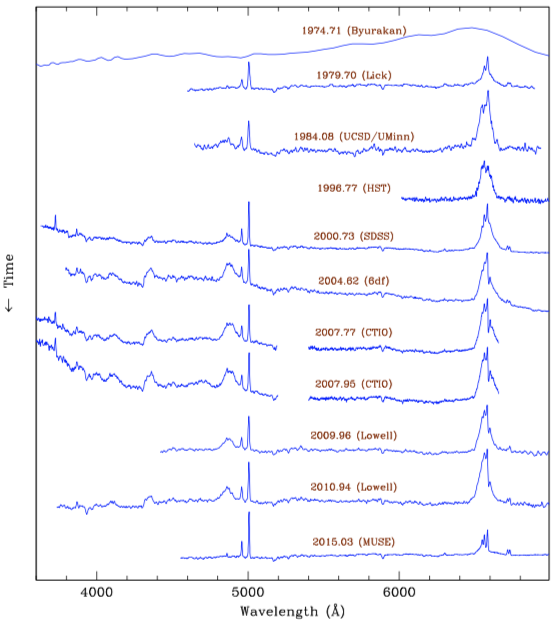}
% \vspace*{-1.0 cm}
    \caption{\textbf{Left:} fading out of the broad emission lines and the featureless blue continuum in the optical spectrum of NGC\,4151 between 1974 and 1984 (lower and upper panels, respectively; \citealt{pen84}). \textbf{Right:} back and forth transition of Mrk\,1018 from a type 2 in late 1979 to a type 1 in 2007 and back to the type 2 class after 2010 \citep{kim18}.}\label{fig_mutants}
\end{center}
\end{figure}

Variability is a well-know characteristic of active nuclei since their discovery \citep[e.g.][]{pet82,cla83}, and has been successfully exploited to infer the size of the innermost components that cannot be spatially resolved with current facilities \citep[e.g.][]{pet98}. The first cases of spectral transitions in Seyfert galaxies were reported very early, even before the unification theory was settled in the field. For instance, the discovery of broad emission lines fading out from the optical spectra of NGC\,4151 and 3C\,390.3 within a timescale of 10 years (left panel in Fig.\,\ref{fig_mutants}) was interpreted by \citet{pen84} as a possible evolutionary sequence in AGN. That is, the presence of the BLR in these nuclei would be linked to the activity of the central continuum source instead of the orientation with respect to the distribution of the obscuring material, thus disappearing when the continuum weakens. Broad emission lines emerging in type 2 and LINERs have also been detected, e.g. in NGC\,1097 \citep{sto93}, NGC\,7582 \citep{are99}, and NGC \,3065 \citep{era01}, usually correlated with variations in the blue-featureless AGN continuum. Furthermore, back and forth transitions have also been detected for some nuclei that have returned to the initial spectral type after experiencing an earlier change, e.g. Mrk\,1018 \citep[][right panel in Fig.\,\ref{fig_mutants}]{kim18}, Mrk\,590 \citep{mat18}, NGC\,1566 \citep{okn19,par19}, NGC\,1365 \citep{ris05}.
Recently, an increasing number of spectral transitions between type 1 and type 2 AGN have been reported, the so-called ``changing-look" quasars \citep[e.g.][]{mat03,lam15}, which can now be identified thanks to the wide sky monitoring surveys such as PanSTARSS \citep{cha16}, CATALINA \citep{dra09}, etc.

In the context of unification such transitions were initially ascribed to the torus structure, and therefore the interpretation was made in terms of variations of the absorption column density. The latter could be caused by e.g. overdensities or clouds in the patchy torus that intercept our line of sight to the central engine, and thus to the BLR that originated the broad lines. However, variable absorption is not compatible with the mid-IR light curves of changing-look AGN, since the expected crossing time for the obscuring material is significantly longer than the observed mid-IR variability \citep{she17}. On the other hand, the low ($\lesssim 1\%$) UV polarisation observed in most of the 13 changing-look quasars observed by \citet{hut19} suggests that such transitions are not likely caused by changes the configuration of the dust obscuring structure. Therefore, it appears that the presence or absence of the BLR may be, at least in part, determined by the activity of the central continuum source, and not by the particular orientation of the system with respect to our line of sight.

\section{The role of evolution and of the host galaxy}

There is evidence accumulated in the last tens of years that nuclear activity is linked to the host galaxy population. This is primarily witnessed by the so-called {\it Magorrian relation} \citep{mag98, fer00} showing a correlation between SMBH masses and velocity dispersion, stellar mass and luminosity of their host galaxies in the local Universe and by the similar shapes, as a function of cosmic time, of the Star Formation (SF) density and BH Accretion (BHA) density, i.e. the so-called {\it AGN/SF co-evolution} \citep{madi14}. Moreover, a possible evolution is envisaged between the various types of active galaxies. Two possible evolutionary progressions are HII $\rightarrow$ Seyfert type 2 \citep{sto01, kau03},
%(Storchi-Bergmann et al 2001; Kauffmann et al 2003), 
or a fuller scenario of HII$\rightarrow$ Seyfert type 2 $\rightarrow$Seyfert type 1 \citep{hun99, lev01, kro02}.
%(Hunt \& Malkan 1999; Levenson et al 2001; Krongold et al 2002). 
These predict that galaxy interactions, leading to the concentration of a large gas mass in the circumnuclear region of a galaxy, trigger starburst emission. Then mergers and bar-induced inflows can bring fuel to a central BH, stimulating AGN activity. While relatively young ($<$1 Gyr) stellar populations are found in more than half of Seyfert 2s \citep{sch99, gon01, rai03}, %(Schmitt et al 1999; Gonz{\'a}lez Delgado et al 2001; Raimann et al 2003), 
they are also found in broad-lined AGN \citep{kau03}. The photometric mid-IR studies of \citet{ede87} and \citet{mai95} did indeed find that Seyfert 2s galaxies more often have enhanced star formation than Seyfert 1s. 

\begin{figure}[t]
% \vspace*{-2.0 cm}
\begin{center}
    \includegraphics[width=6.5cm]{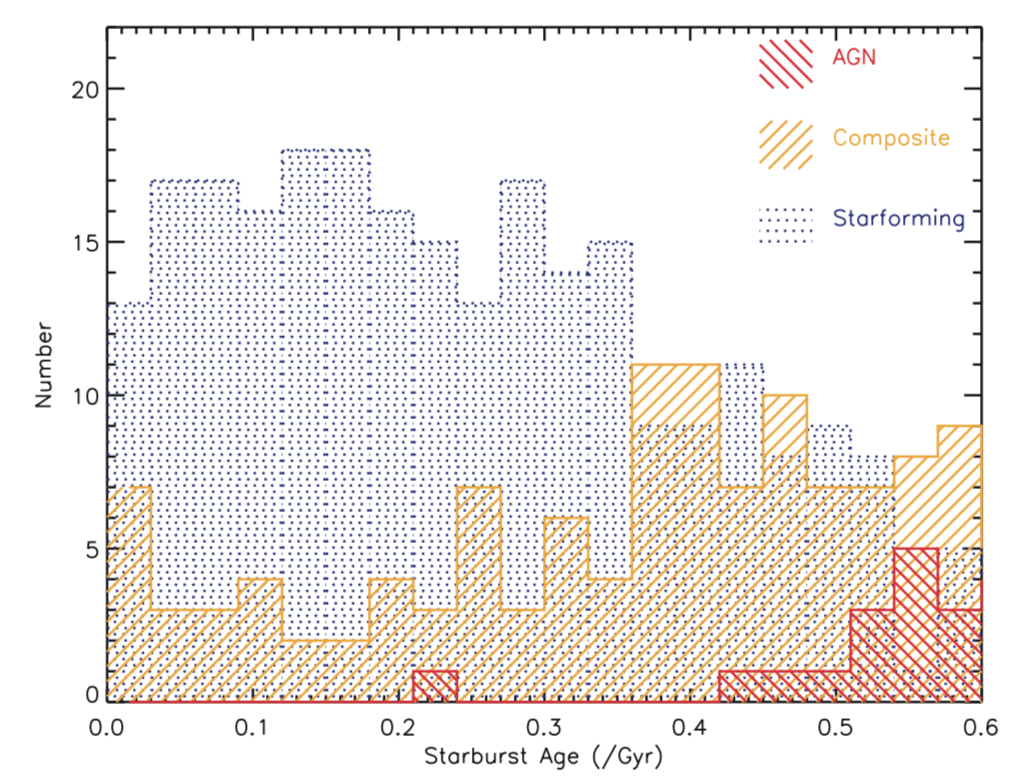}~\includegraphics[width=7.2cm]{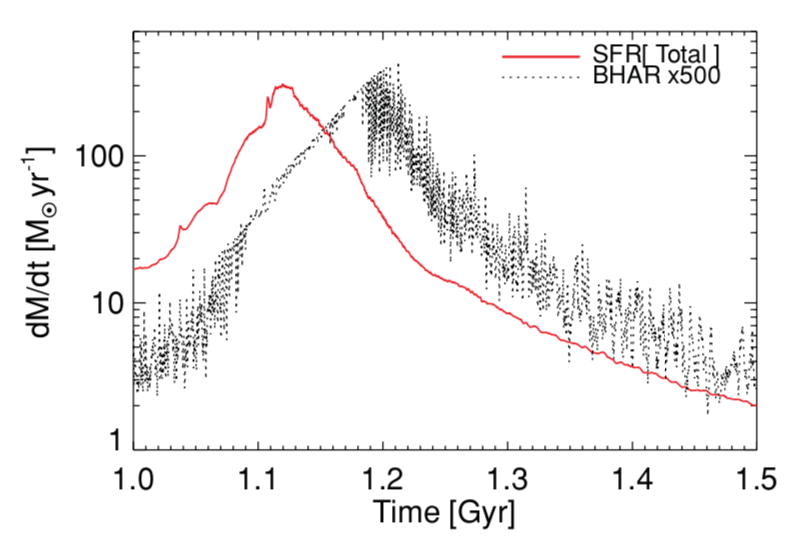}
% \vspace*{-1.0 cm}
    \caption{{\bf Left: (a)}  the number of star-forming (blue), composite-AGN (orange) and pure-AGN (red) in our sample as a function of time since the onset of the starburst. Figure from \citet{wil10}. {\bf Right: (b)} Galaxy merger simulation, near coalescence of the two galaxy nuclei (at t $\sim$ 1.1 Gyr). The total SFR peaks near this coalescence, as inflows first reach $\sim$ kpc scales. The BHAR (here multiplied by 500 for ease of comparison) grows rapidly during this time, but sufficient gas remains to fuel BH growth until $\sim$ 10$^8$ yr later. Figure from \citet{hop12}. }\label{fig_evol}
\end{center}
\end{figure}

A temporal link, if not a casual link, between SF and BHA has been demonstrated both observationally and theoretically. \citet{wil10} studied the growth of BHs, with masses of $10^{6.5} - 10^{7.5}\, \rm{M_\odot}$ and AGN luminosities of $10^{42} - 10^{44}\, \rm{erg\,s^{-1}}$, in 400 local galactic bulges which have experienced a strong burst of star formation in the past 600\,Myr. During the first 600\,Myr after a starburst, the BHs in the sample increase their mass by 5\% on-average and the total mass of stars formed is about $10^3$ times the total mass accreted onto the BH. This ratio is similar to the ratio of stellar to BH mass observed in present-day bulges. They also find that the average rate of accretion of matter onto the BH rises steeply roughly 250\,Myr after the onset of the starburst (Fig.\,\ref{fig_evol}-a). \citet{hop12} simulations of AGN fueling by gravitational instabilities naturally produce a delay between the time when SFRs peak inside of a given annulus and the time when AGN activity peaks (Fig.\,\ref{fig_evol}-b). This offset scales as the gas consumption time, $\sim 10-100$ dynamical times. On small scales ($\lesssim 10\, \rm{pc}$), this is characteristically $\sim 10^7\, \rm{yr}$, rising to a few  $10^8\, \rm{yr}$ on kpc scales. These offsets are similar to the magnitude of time offsets suggested by various observations on both small and large scales \citep{wil10}.

Furthermore, the evolution of AGN and their observational appearance is connected to the evolution of the host galaxy \citep{bal06}. The ratio of obscured to unobscured AGN (R) increases with redshift, implying a change to the traditional unification model. Since the obscuring medium is changing with redshift, it must be influenced, e.g., by the cosmic star formation rate, which peaks at a very similar redshift as R. As star formation increases in a galaxy, also the absorbing gas and dust increase, acting as an obscuring medium. The absorbing material would be located either close to the dust sublimation radius (and mimic some properties of the absorbing torus), or it could also be spread over most of the inner part of the galaxy \citep{mcl95}.
%(McLeod \& Rieke 1995). 
The idea of an extended, more galactic-scale obscuring medium is consistent with IR observations, which have pointed out the remarkable similarity in the mid-to-far-IR emission between Seyfert 2s and 1s \citep{kur03,lut04}, 
%(Kuraszkiewicz et al. 2003; Lutz et al. 2004), 
in contrast to the predictions of the simple molecular torus model. 
\citet{bal06} conclude that most of the accretion in the universe is obscured and that this obscuration evolves similar to the star formation rate. These facts deepen the connection between star formation and AGN fueling, as well as that between BH growth and galaxy evolution.

\citet{buc06}, using the 12$\mu$m Seyfert galaxies sample \citep{spi89,rus93}, find that the Seyfert 2 galaxies typically show stronger starburst contributions than the Seyfert 1 galaxies in the sample, contrary to what is expected based on the unified scheme for AGN. \citet{tom10}, using the same local sample of AGN, found that the mid-IR emission properties characterize all the AGN 1’s (which include both Seyfert type 1 and Hidden Broad Line galaxies) as a single family, with strongly AGN-dominated spectra. In contrast, the AGN 2’s can be divided into two groups, the first one with properties similar to the AGN 1’s except without detected broad lines, and the second with properties similar to the non-Seyfert galaxies, such as LINERs or starburst galaxies.

\begin{figure}[t]
% \vspace*{-2.0 cm}
\begin{center}
    \includegraphics[width=14cm]{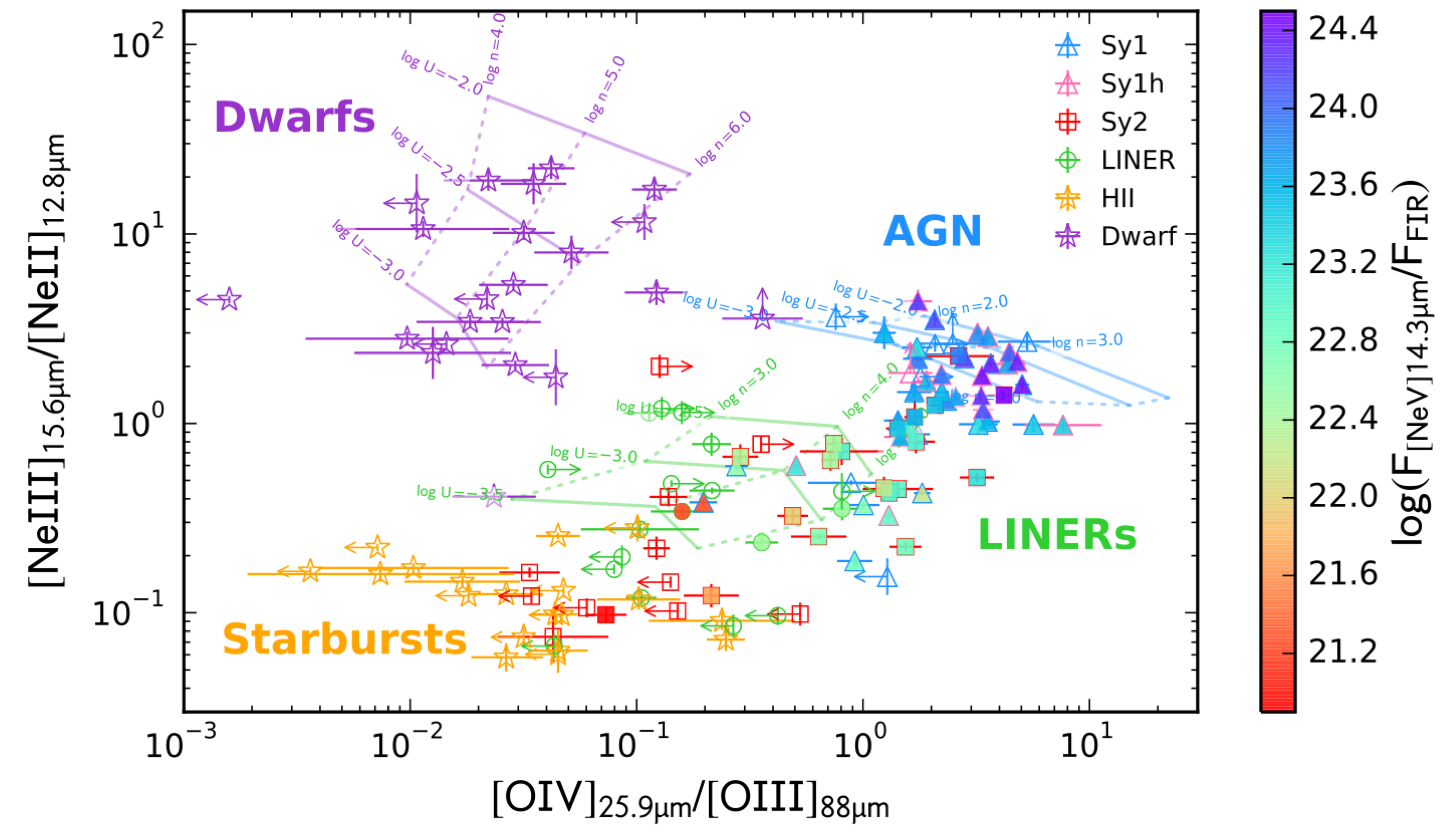}
% \vspace*{-1.0 cm}
    \caption{The [Ne\,\textsc{iii}]$_{\rm 15.6 \mu m}$/[Ne\,\textsc{ii}]$_{\rm 12.8 \mu m}$ line ratio {\it vs} the [\textsc{O\,iv}]$_{\rm 25.9 \mu m}$/[\textsc{O\,iii}]$_{\rm 88 \mu m}$ ratio for AGN with different classifications based on the optical spectrum. Photoionisation models of AGN, LINER, starburst galaxies, and dwarf galaxies are shown as blue, green, yellow, and purple grids, respectively. The logarithmic values of the density ($n_{\rm H}$) and ionisation potential ($U$) of the photoionisation models are indicated in the figures. Symbols are colour-coded according to their $\rm F_{\rm [Ne\,V]14.3 \mu m}/F_{\rm FIR}$ flux ratio, when available (see colour bar). Figure from \citet{fer16}.}\label{fig_ir_bpt}
\end{center}
\end{figure}

The combination of {\it Spitzer} and {\it Herschel} mid- and far-IR spectroscopy, respectively, of Seyfert galaxies, LINER and dwarf galaxies in the Local Universe, allowed us to define a line ratio diagram, the so-called {\it new IR BPT diagram} which can separate the various types of AGN through the ionized fine-structure lines \citep{fer16}. We report such diagram in Fig.\,\ref{fig_ir_bpt}, where we can easily see that the position of Seyfert type 1 galaxies is well displaced from that one of the Seyfert type 2 galaxies. Many of these latter galaxies lie closer to the Starburst galaxies region, while the LINER occupy a region which is intermediate between the Starburst galaxies and the Seyfert type 1 galaxies. We confirm here the result of \citet{tom10} that the ``hidden broad-line region galaxies are indistinguishable form the Seyfert type 1's''.

\section{Conclusions}

Our conclusions can be summarized as follows:
\begin{itemize}
    \item The ``unified model" has been conceived to explain the large {\it zoo} of different AGN with a single physical object, however, while it is well recognized that it works for unificating the so-called ``Hidden Broad-Line Region Galaxies" (HBLR), i.e. those galaxies for which, either polarization spectra or near-IR ones, show broad lines, with the AGN type 1, it cannot be generalized to all AGN types.
\item It seems so far that no strong observational evidence of tori with the needed characteristics to block BLR, collimate radiation and let AGN feeding has yet been found.
\item In a large fraction of the changing-looking AGN, ``transitions" can be explained by the AGN duty cycle.
\item AGN and host galaxy interrelations may indicate an evolutionary path, e.g. HII $\rightarrow$ AGN2 $\rightarrow$ AGN1.
\item In the next years the models of unification need to also consider this observational framework and not only simple orientation effects. 
\end{itemize}

\vspace{-0.4cm}
\section*{Acknowledgements}
\textit{LS acknowledges Mirjana Povic for her kind invitation at the conferece in Addis Ababa, organized with the Ethiopian Space Science and Technology Institute (ESSTI). LS and JAFO acknowledge financial support by the Agenzia Spaziale Italiana (ASI) under the research contract 2018-31-HH.0.}

\end{document}